\documentstyle [12pt]{article}
\def\int {\intop \limits}

\def\fnote#1{\footnote}
\begin{document}
\newcommand{\dst}[1]{\displaystyle{#1}}
\newcommand{\barl}{\begin{array}{rl}}
\newcommand{\ball}{\begin{array}{ll}}
\newcommand{\ear}{\end{array}}
\newcommand{\barc}{\begin{array}{c}}
\newcommand{\sne}[1]{\displaystyle{\sum _{#1} }}
\newcommand{\sn}[1]{\displaystyle{\sum ^{\infty }_{#1} }}
\newcommand{\ini}[1]{\displaystyle{\int ^{\infty }_{#1}}}
\newcommand{\myi}[2]{\displaystyle{\int ^{#1}_{#2}}}
\newcommand{\inn}{\displaystyle{\int }}
\newcommand{\be}{\begin{equation}}
\newcommand{\ee}{\end{equation}}
\newcommand{\aq}[1]{\label{#1}}
\renewcommand \theequation{\thesection.\arabic{equation}}

\vspace*{4.0cm}
\centerline{\Large {\bf Multi-photon effects in energy}}
\vskip .25cm
\centerline{\Large {\bf losses spectra}}
\vskip .5cm
\centerline{\large{\bf V. N. Baier and V. M. Katkov}}
\centerline{Budker Institute of Nuclear Physics,
 630090 Novosibirsk, Russia}
\vskip 2.0cm
\begin{abstract}
Effect of radiation of many photons by a single electron traversing
a target is discussed. When the summary energy of emitted photons
(the energy losses spectrum) is measured only, the photon spectrum
is distorted comparing with the photon spectrum in one interaction.
Influence of this effect is discussed for the cases (1) bremsstrahlung
(described by Bethe-Heitler formula), (2)
the strong Landau-Pomeranchuk-Migdal effect
and (3) transition radiation.
Qualitative picture of the phenomenon is discussed in detail.
Comparison with the recent SLAC experiment in relatively thick target
(2.7 \% of the radiation length), where the effect of
emission of many photons by a projectile is very essential,
shows perfect agreement of the theory and data.
\end{abstract}

\newpage
\section{Introduction}

A very successful series of experiments \cite{5} - \cite{7} was performed
at SLAC during last years. In these experiments the cross section
of bremsstrahlung of soft photons with energy from 200~KeV to
500~MeV from electrons with energy 8~GeV and 25~GeV is measured
with an accuracy of the order of a few percent. Both LPM
(the Landau-Pomeranchuk-Migdal effect, i.e. suppression of the
bremsstrahlung from high energy electron
due to a multiple scattering of an emitting electron in dense media
\cite{1}- \cite{3})
and dielectric suppression
(the effects of the polarization of a medium, \cite{4})
are observed and investigated. These experiments were the
challenge for the theory since in all the previous papers calculations
(cited in \cite{8}) are performed to
logarithmic accuracy which is not enough for description
of the new experiment. The contribution of the Coulomb corrections (at least
for heavy elements) is larger then experimental errors and these corrections
should be taken into account.

Very recently authors developed the new approach to the theory of LPM
effect \cite{8},
where the cross section of bremsstrahlung process
in the photon energies region where the influence of the LPM is very strong
was calculated with term $\propto 1/L$ , where $L$
is characteristic logarithm of the problem,
and with the Coulomb corrections
taken into account. In the photon energy region, where the LPM effect
is "turned off", the obtained cross section
gives the exact Bethe-Heitler cross section (within power accuracy) with
Coulomb corrections. This important feature was absent in
the previous calculations. The contribution of an inelastic scattering of
a projectile on atomic electrons is also included.
The polarization of a medium is incorporated into
this approach. The considerable contribution into the soft part of
the measured spectrum of radiation gives a photon
emission on the boundaries of a target. We calculated this contribution
taking into account the multiple scattering and polarization of a medium
for the case when a target is much thicker than the
formation length of the radiation. We considered also
a case when a target is much thinner than the
formation length. A case of an intermediate thickness of a target (between
cases of a thick and a thin target) is analyzed later in \cite{8b}.

It should be noted that in the experiments \cite{5}-\cite{7}
the summary energy of all photons radiated by a single electron
is measured. This means that besides mentioned above
effects there is an additional "calorimetric" effect due to the multiple
photon radiation. This effect is especially important in relatively
thick used targets. Since the energy losses spectrum of an electron
is actually measured, which is not coincide in this case with the
spectrum of photons radiated in a single interaction,
one have to consider the distribution function of electrons
over energy after passage of a target. As it is known,
this distribution function is the solution of the corresponding kinetic
equation. However, the problem can be simplified essentially if one is
interested in the soft part of the energy losses spectrum.
Just this situation was in the experiments \cite{5}-\cite{7},
since the measurements were performed in the region
of the photon energies of one to five orders of magnitude lower than the
electron energy.

In Sec. 2 of the present paper the general approach to the
consideration of the soft part of the energy losses spectrum due to
of many photons emission from a single electron is 
formulated without limitation of the target thickness.
The case when the formation length is larger than the target thickness is
analyzed also.
Using this approach we calculated corrections (in a form of 
the reduction factor) to the spectrum of energy losses for
the Bethe-Heitler case as well as for the situation when the LPM effect
is strong and when the transition radiation is essential. In Sec. 3 we
analyze basic properties of the phenomenon and apply the theory developed
in \cite{8}  with multiphoton corrections
to discussion of data \cite{7}. We conclude that the theory is in a perfect
agreement with data. In Appendix we calculate the radiation probability 
of arbitrary number of hard photons.

\section{Spectral distribution of the energy losses}
\setcounter{equation}{0}

The probability of the successive radiation of $n$ soft photons
with energies \newline $\omega_1, \omega_2, \ldots \omega_n$
by a particle with energy $\varepsilon~(\omega_k \ll \varepsilon, k=1,2
\dots n)$ on the length $l$
in the energy intervals $d\omega_1 d\omega_2 \ldots d\omega_n$ is given
by expression (in this paper, we employ units such that $\hbar=c=1$)
\begin{eqnarray}
&& dw(\omega_1, \omega_2, \ldots \omega_n)=
A\int_{0}^{l}dW(\omega_1)dl_1 \int_{0}^{l_1}dW(\omega_2)dl_2 \ldots
\int_{0}^{l_{n-1}}dW(\omega_n)dl_n \nonumber \\
&&=\frac{A}{n!} dw(\omega_1)dw(\omega_2) \ldots dw(\omega_n),
\label{1}
\end{eqnarray}
where $A$ is the normalization constant, $dW(\omega)/d\omega$ is
the differential probability of the photon radiation per unit length,
$dw(\omega)/d\omega=ldW(\omega)/d\omega$ is the differential
probability of the photon radiation per length $l$. If the probability
$dw/d\omega$ doesn't depend on the particle energy $\varepsilon$ then
integrating (\ref{1}) over all photon energies we obtain
\begin{equation}
w_n=\frac{A}{n!}w^n,\quad w=\int_{}^{}\frac{dw}{d\omega}d\omega.
\label{2}
\end{equation}
The value $A$ is defined by the condition that probability of all
the possible events with radiation of any number of photons or without
photon radiation is equal to unit.
\begin{equation}
\sum_{n=0}^{\infty}w_n=A\sum_{n=0}^{\infty}\frac{1}{n!}w^n=A\exp w=1,\quad
A=\exp(-w).
\label{3}
\end{equation}
Using (\ref{1}),(\ref{3}) we can write the expression for the
differential distribution of the energy losses in the form
\begin{equation}
\frac{1}{\omega}\frac{d\varepsilon}{d\omega}=
\sum_{n=1}^{\infty}\frac{1}{n!}\exp (-w)\int_{}^{}\frac{dw}{d\omega_1}
\frac{dw}{d\omega_2} \cdots \frac{dw}{d\omega_n}
\delta\left(\sum_{k=1}^{n}\omega_k-\omega\right)d\omega_1 \ldots d\omega_n.
\label{4}
\end{equation}
Here on on the right-hand side we have the sum of the probabilities
of radiation of $n$ photons with summary energy $\omega$.

Using standard parametrization of $\delta$-function
\[
\delta\left(\sum_{k=1}^{n}\omega_k-\omega\right)
=\frac{1}{2\pi}\int_{-\infty}^{\infty}
\exp \left(is\left(\omega-\sum_{k=1}^{n}\omega_k\right)\right)ds
\]
we obtain
\begin{eqnarray}
&& \frac{d\varepsilon}{d\omega}=\frac{\omega}{2\pi}
\int_{-\infty}^{\infty}\exp\left(is\omega-w \right)
\sum_{1}^{\infty}\frac{1}{n!}\left(\int_{}^{}\frac{dw}{d\omega_1}
\exp (-is\omega_1)d\omega_1 \right)^n ds \nonumber \\
&& =\frac{\omega}{2\pi}
\int_{-\infty}^{\infty}\exp\left(is\omega \right)
\exp\left\{-\int_{0}^{\infty}\frac{dw}{d\omega_1}
\left[1-\exp(-is\omega_1) \right] d\omega_1\right\}ds \nonumber \\
&& =\frac{1}{\pi}~{\rm Re} \int_{0}^{\infty}\exp\left(ix \right)
\exp\left\{-\int_{0}^{\infty}\frac{dw}{d\omega_1}
\left[1-\exp\left(-ix\frac{\omega_1}{\omega}\right)\right]
d\omega_1\right\}dx
\label{5}
\end{eqnarray}
The formula (\ref{5}) was derived by Landau \cite{9a} as solution
of the kinetic equation under assumption that energy losses are
much smaller than particle's energy (the paper \cite{9a} was devoted
to the ionization losses). Let us notice the following.
The energy losses are defined by the hard part of the radiation spectrum.
In the soft part of the energy losses spectrum (\ref{5})
the probability of radiation of one hard photon only is taken into account
accurately. To calculate the probability of the emission of two and more
hard photons one has to take into account step by step the recoil
in previous acts of the photon emission. 
The probability of radiation of two and more hard photons is of the order
$(l/L_{rad})^2$ and so on. Thus, the formula (\ref{5}) is applicable for the
thin targets and has accuracy $l/L_{rad}$. If we want to improve accuracy of
the formula (\ref{5}) and for the case of thick targets $l \geq L_{rad}$
one has to consider radiation of an arbitrary number of hard photons.
This problem is solved in Appendix for the case when hard part of
the radiation spectrum is described by the Bethe-Heitler formula.
In this case the formula (\ref{5}) acquires the additional factor. As a
result we extend this formula on the case thick targets.

We will analyze first the interval of photon energies where
Bethe-Heitler formula is valid. We write it in the form
\begin{equation}
\omega \frac{dw}{d\omega}=\frac{l}{L_{rad}}\left[\frac{4}{3}
\left(1-\frac{\omega}{\varepsilon} \right)+\frac{\omega^2}{\varepsilon^2}
\right] \vartheta(\varepsilon-\omega)=
\beta\left(1-x_1+ \frac{3}{4}x_1^2 \right)\vartheta(1-x_1),
\label{7}
\end{equation}
where $l$ is the thickness of the target, $L_{rad}$ is the radiation
length, 
\[
\beta=\frac{4l}{3L_{rad}},
\quad x_1=\frac{\omega}{\varepsilon}.
\]
The integral in the curly brackets in (\ref{5}) we split into
two integrals with integration intervals
$(0, \omega')$ and $(\omega', \varepsilon)$,
where we choose $\omega'$ such that $\omega \ll \omega' \ll \varepsilon$
(note, that the radiation probability vanishes for $\omega > \varepsilon$).

We will begin with consideration of rather thin
targets such that $l \ll L_{rad}$
(the case $l \sim L_{rad}$ will be discussed below). In this case
in the integral over $x$ in (\ref{5}) contributes mainly $x \sim 1$ and
the integral over the second interval is
\begin{equation}
J_2 \simeq \int_{\omega'}^{\varepsilon}\frac{dw}{d\omega_1}
\left[1-\exp\left(-ix\frac{\omega_1}{\omega}\right) \right] d\omega_1
\simeq \int_{\omega'}^{\varepsilon}\frac{dw}{d\omega_1}d\omega_1
=\beta \left(\ln \frac{\varepsilon}{\omega'} -\frac{5}{8} \right), 
\label{8}
\end{equation}
The integral containing $\displaystyle{\exp\left(-ix
\frac{\omega_1}{\omega}\right)}$ vanishes
because of the fast oscillations of the integrand.
The integration over the first interval gives
\begin{equation}
 J_1 = \int_{0}^{\omega'}\frac{dw}{d\omega_1}
\left[1-\exp\left(-ix\frac{\omega_1}{\omega}\right) \right]d\omega_1
\simeq \beta \left[\ln \left(\frac{\omega'}{\omega}x \right)+C+
i\frac{\pi}{2} \right],
\label{9}
\end{equation}
where $C=0.577 \ldots$ is the Euler constant.
So, the total integral over $\omega_1$ in (\ref{5}) is
\begin{equation}
J=J_1+J_2 \simeq \beta\left(\ln x + \mu + i\frac{\pi}{2} \right),\quad
\mu=\ln \frac{\varepsilon}{\omega}-\frac{5}{8}+C.
\label{10}
\end{equation}
Substituting this result into (\ref{5}) we have
\begin{equation}
\frac{d\varepsilon}{d\omega}=\frac{1}{\pi}{\rm Re}\int_{0}^{\infty}
\exp \left[ix-\beta\left(\ln x + \mu + i\frac{\pi}{2} \right) \right] dx
\label{11}
\end{equation}
In this integral we turn the integration contour over $x$ by angle
$\frac{\pi}{2}$ and substitute $x \rightarrow ix$. As a result we find
\begin{eqnarray}
&& \frac{d\varepsilon}{d\omega}=\frac{1}{\pi}{\rm Re}~i\exp (-i\beta \pi)
\exp (-\beta \mu)\int_{0}^{\infty} x^{-\beta}e^{-x}dx \nonumber \\
&& =\beta \exp (-\beta \mu) \Gamma(1-\beta) \frac{\sin \beta \pi}{\beta \pi}
=\beta \frac{\exp (-\beta \mu)}{\Gamma(1+\beta)}
\label{12}
\end{eqnarray}
where $\Gamma(z)$ is the Euler gamma function.
If we consider radiation of the one soft photon, we have from
(\ref{5}) $d\varepsilon/d\omega=\beta$. Thus, formula (\ref{12})
gives additional "reduction factor" $f_{BH}$ which
characterizes the distortion
of the soft Bethe-Heitler spectrum due to multiple photons radiation.
With regard for factor $g(\beta)$ (see Appendix Eqs.(\ref{a.15}),
(\ref{a.16}) we obtain the reduction factor valid for $\beta \geq 1$
while Eq.(\ref{12}) is true for $\beta \ll 1$)
\begin{equation}
f_{BH}=g(\beta)\frac{\exp (-\beta \mu)}{\Gamma(1+\beta)}
=\left(\frac{\omega}{\varepsilon}\right)^{\beta}(1+\beta)^{1/4}
\left(1+\frac{\beta}{2}\right)^{3/4}.
\label{13}
\end{equation}
In Fig.1 the function $f_{BH}(\omega)$ is given for electron with
the energy $\varepsilon$ = 25~GeV. One can see that for $\omega$ = 100~MeV
the value $f_{BH}(\omega)$ differs from unit in 5-6 times greater than value
$\beta$. Let us discuss this circumstance. The emission of accompanying
photons with energy much lesser or of the order of $\omega$
(we consider in this figure the situation connected with SLAC experiment
\cite{5}-\cite{7} where $\omega \ll \varepsilon,~\beta
\ll 1 $) changes the spectral distribution on quantity order of $\beta$.
However, if one photon with energy $\omega_r > \omega$
is emitted, at least, then photon with energy $\omega$ is not registered
at all in the corresponding chanel of the calorimeter. Since mean number
of photons with energy larger than $\omega$ is determined by the
expression (see (\ref{1})-(\ref{5}))
\begin{equation}
\sum_{n=0}^{\infty} n w_n =w_{\omega}=\int_{\omega}^{\varepsilon}
\frac{dw}{d\omega}d\omega,
\label{13a}
\end{equation}
the probability of the event when no photon with energy $\omega_r > \omega$ 
is radiated is defined by
$\exp \left(-w_{\omega} \right)$. This is just the main factor in
the expression for the reduction factor $f_{BH}$ (\ref{13}) while the
difference begins with the terms $\sim \beta^2$. In the case, when
radiation is described by the Bethe-Heitler formula the value
$w_{\omega}$ increases as a logarithm with $\omega$ decrease
($w_{\omega} \simeq \beta \ln \varepsilon/\omega$) and for large
ratio $\varepsilon/\omega$ the value $w_{\omega}$ is much larger than $\beta$.
Thus, amplification of the effect is connected with large interval of
the integration at evaluation of the radiation probability of a hard photon.

The Bethe-Heitler formula becomes inapplicable for the photon energies
$\omega \leq \omega_c$, where LPM effect starts to manifest itself
(see \cite{8})
\begin{equation}
\omega_c=\frac{4\pi}{\alpha}\frac{\gamma^2}{L_{rad}},
\label{14}
\end{equation}
where $\alpha=1/137,~\gamma=\varepsilon/m$. The expression for the photon
spectrum for this case can be obtained from \cite{8}, Sec.2
\begin{eqnarray}
&& \omega\frac{dw}{d\omega}=\beta \chi(\nu_0),\quad
\chi(\nu_0)=3~{\rm Im}~\int_{0}^{\infty} \exp (-it) \left(
\frac{1}{\sinh^2 z}-\frac{1}{z^2} \right) dt; \nonumber \\
&& z=\nu t,\quad \nu^2=i\nu_0^2,\quad \nu_0^2=\frac{\omega_c}{\omega}.
\label{15}
\end{eqnarray}
In this expression the terms $\sim 1/L$ (see \cite{8}) are not taken
into account, since the terms of this type can be neglected at
calculation of the reduction factor (the terms of the order $\beta/L$).
Using (\ref{15}) we calculate now the total probability of radiation $w$.
As it was done above, we divide the integration interval into two parts
$(0, \omega')$ and $(\omega', \varepsilon)$ where
$\omega_c \ll \omega' \ll \varepsilon$. In the interval
$(\omega', \varepsilon)$ the Bethe-Heitler formula is valid and we
have for $w_1$ formula (\ref{8}). The integral over the second interval is
\begin{eqnarray}
&&\hspace{-18mm} w_2=\int_{0}^{\omega'}\frac{dw}{d\omega}d\omega=3\beta~
{\rm Im}~\int_{0}^{\omega'}\frac{d\omega}{\omega} \int_{0}^{\infty}
\exp (-it) \left(\frac{1}{\sinh^2 z}-\frac{1}{z^2} \right) dt
\nonumber \\
&&\hspace{-18mm} = 6\beta~{\rm Im}~\int_{0}^{\infty}\Phi(z_0) \exp (-it)dt;~
\Phi(z_0)=\int_{z_0}^{\infty}\frac{dz}{z}
\left(\frac{1}{\sinh^2 z}-\frac{1}{z^2} \right),~
z_0^2=it^2\frac{\omega_c}{\omega'}.
\label{16}
\end{eqnarray}
Performing integration by parts in the expression for $\Phi(z_0)$ we have
\begin{equation}
\Phi(z_0)=\ln z_0 \left(\frac{1}{z_0^2}-\frac{1}{\sinh^2 z_0} \right)
+2\int_{z_0}^{\infty}\ln z\left(\frac{\cosh z}{\sinh^3 z}-
\frac{1}{z^3} \right)dz.
\label{17}
\end{equation}
Into the integral over $t$ in (\ref{16}) the interval $t \leq 1$
contributes, so that $|z_0| \ll 1$. Because of this we can expand the
first term on the right-hand side of formula (\ref{17})
and  put the lower limit of integration equal to zero
in the second term and than turn the integration
contour by angle $-\pi/4$. We find
\begin{equation}
\Phi(z_0)=\frac{\ln z_0}{3}
+2\int_{0}^{\infty}\ln z\left(\frac{\cosh z}{\sinh^3 z}-
\frac{1}{z^3} \right) dz.
\label{18}
\end{equation}
Using the integrals
\begin{equation}
\int_{0}^{\infty}\exp (-it)dt=-i,\quad \int_{0}^{\infty}\ln z_0 \exp (-it)dt
=i\left(\frac{1}{2}\ln \frac{\omega'}{\omega_c}+C \right),
\label{19}
\end{equation}
we have
\begin{equation}
w_2=\beta\left(\ln \frac{\omega'}{\omega_c}+2C +C_1 \right),\quad
C_1=12 \int_{0}^{\infty}\ln z\left(\frac{1}{z^3} - \frac{\cosh z}{\sinh^3 z}
 \right) dz \simeq 1.4294.
\label{20}
\end{equation}
Adding to this result the probability of radiation $w_1$ we obtain
for the total probability of radiation
\begin{equation}
w_c=w_1+w_2=\beta\left(\ln \frac{\varepsilon}{\omega_c}+C_2 \right),\quad
C_2=2C+C_1-\frac{5}{8} \simeq 1.959.
\label{21}
\end{equation}

We consider now the spectral distribution of the energy losses
in the conditions of the strong LPM effect $(\omega \ll \omega_c)$.
In this case the contribution into the integral defining the function
$\chi(\nu_0)$ in (\ref{15}) gives an interval $|z| \leq 1,
~t \leq 1/\nu_0=\sqrt{\omega/\omega_c} \ll 1$, so that $\exp (-it) \simeq 1$.
We have as a result
\begin{equation}
\chi(\nu_0)=-3~{\rm Im}~\frac{1}{\nu}=\frac{3}{\sqrt{2} \nu_0};\quad
\frac{dw}{d\omega}=\frac{3}{\sqrt{2}}
\frac{\beta}{\sqrt{\omega \omega_c}}
\label{22}
\end{equation}
Let us use this result in the integral in formula (\ref{5})
\begin{eqnarray}
&& \int_{0}^{\infty}\frac{dw}{d\omega_1}\exp
\left(-ix\frac{\omega_1}{\omega}\right)
d\omega_1 = \frac{3}{\sqrt{2}} \beta \int_{0}^{\infty}
\frac{d\omega_1}{\sqrt{\omega_1 \omega_c}}
\exp \left(-ix\frac{\omega_1}{\omega}\right) \nonumber \\
&& =3\beta\sqrt{\frac{\pi \omega}{2x\omega_c}}
\exp \left(-i\frac{\pi}{4}\right) \equiv \varphi(x)
\label{23}
\end{eqnarray}
Substituting (\ref{21}) and (\ref{23}) into formula (\ref{5}) and
taking into account the factor $g(\beta)$ (\ref{a.15})
we obtain for the energy losses
\begin{equation}
\frac{d\varepsilon}{d\omega}=\exp (-w_c)\frac{g(\beta)}{\pi}~{\rm Re}~
\int_{0}^{\infty} \exp \left(ix+\varphi(x)\right) dx
\label{24}
\end{equation}
The main contribution to the last integral gives $x \sim 1$.
Then $|\varphi(x)| \ll 1$ and one can expand the integrand in
(\ref{24}) in powers of $\varphi(x)$. Conserving the two first term
of the expansion we have integrals
\begin{equation}
{\rm Re}~\exp \left(-i\frac{\pi}{4}\right)
\int_{0}^{\infty}\frac{dx}{\sqrt{x}}
\exp (ix)=\sqrt{\pi},\quad {\rm Re}~\exp \left(-i\frac{\pi}{2}\right)
\int_{0}^{\infty}\frac{dx}{x} \exp (ix)=\frac{\pi}{2}
\label{25}
\end{equation}
Using this integrals we find for the distribution of the spectral
energy losses
\begin{equation}
\frac{d\varepsilon}{d\omega}=3\beta \sqrt{\frac{\omega}{2\omega_c}}f_{LPM},
\quad f_{LPM} = g(\beta)\left(1+\frac{3\pi}{2\sqrt{2}}\beta
\sqrt{\frac{\omega}{\omega_c}} \right) \exp (-w_c),
\label{26}
\end{equation}
where $f_{LPM}$ is the reduction factor in the photon energy range
where the LPM effect is essential.
In Fig.2 the function $f_{LPM}(\omega)$ is shown for $\omega <$100~MeV
($\varepsilon$ = 25~GeV, $\omega_c$ = 228~MeV). Taking in consideration that
$\omega \ll \omega_c$ and using Eq.(\ref{22})  we find (see also
discussion after (\ref{13}))
\begin{eqnarray}
&& w_{\omega}=\int_{\omega}^{\varepsilon}\frac{dw}{d\omega}d\omega
=w_c-\int_{0}^{\omega}\frac{dw}{d\omega}d\omega
\simeq w_c-3\sqrt{2}\beta \sqrt{\frac{\omega}{\omega_c}} \nonumber \\
&& \exp \left(-w_{\omega} \right) \simeq \left(1+
3\sqrt{2}\beta \sqrt{\frac{\omega}{\omega_c}} \right)
\exp \left(-w_{c} \right)
\label{26a}
\end{eqnarray}
In the last expression the coefficient 3$\sqrt{2}$
entering into the term $\sim \beta$ differs from exact value in (\ref{26})
on factor $4/\pi$=1.27 only. It is seen in Fig.2 that the reduction
factor $f_{LPM}$ changes appreciably in the region of high photon energies
solely. This is due to the fact that here the total probability of
photon radiation is finite (in contract to the Bethe-Heitler formula)
and the integral which defines this probability converges at
$\omega \rightarrow 0$.

In the above analysis we neglected an influence of the
polarization of a medium on the bremsstrahlung. This is correct if
$\kappa \ll \nu_0$ (see (\ref{15}) and \cite{8}, Sec.3) where
\begin{equation}
\kappa=1+\frac{\omega_p^2}{\omega^2} \equiv 1+ \kappa_0^2,\quad
\omega_p=\gamma \omega_0,\quad \omega_0^2=\frac{4\pi \alpha n}{m},
\label{27}
\end{equation}
where $n$ is the electron density.
We assume, for definiteness, that $\omega_p \ll \omega_c$. This is true
in any case for a dense matter if electron energy $\varepsilon \geq 10~GeV$.
In the opposite case when $\kappa \gg \nu_0 \gg 1$ there is an additional
suppression of the bremsstrahlung (see \cite{8}, Sec.3):
\begin{equation}
\omega \frac{dw}{d\omega} \simeq \frac{3l}{4\kappa L_{rad}}
= \frac{\beta}{\kappa}
\label{28}
\end{equation}
and this contribution into reduction factor (\ref{26}) can be neglected.

The main contribution into effect considered for the photon
energies such that $\omega \ll \omega_p$ gives the transition radiation
The differential probability of the transition radiation on the boundary
of the vacuum and a medium is (neglecting interference between nearby
edges, see, e.g. \cite{9b}, Sec.14)
\begin{equation}
\omega \frac{dw_{tr}}{d\omega}=\frac{2\alpha}{\pi}\left[\left(1+
\frac{2}{\kappa-1} \right)\ln \kappa -2 \right] \equiv \eta J_{tr},
\label{29}
\end{equation}
where notations are introduced
\begin{equation}
\eta=\frac{2\alpha}{\pi},\quad J_{tr}=\left(1+
\frac{2}{\kappa-1} \right)\ln \kappa -2.
\label{30}
\end{equation}
Let us note, that for the transition radiation the given above derivation
of the expression (\ref{5}) is directly inapplicable. However,
for application of formula (\ref{5}) it is enough that acts of radiation
of soft photons are statistically independent.
We calculate now the integral over $\omega_1$ in (\ref{5}) splitting as above
the integration interval over $\omega_1$ into two parts: $(0,\omega')$ and
$(\omega', \infty)$ where $\omega \ll \omega' \ll \omega_p$.
In the upper interval we have
\begin{equation}
\int_{\omega'}^{\infty} J_{tr}(\omega_1)
\left[1-\exp\left(-ix\frac{\omega_1}{\omega}\right) \right]
\frac{d\omega_1}{\omega_1} \simeq
\int_{\omega'}^{\infty} J_{tr}(\omega_1) \frac{d\omega_1}{\omega_1}
\simeq \left(\ln \frac{\omega_p}{\omega'}-1 \right)^2 +\frac{\pi^2}{12}.
\label{31}
\end{equation}
In the second interval ($\omega_1 \leq \omega'$) we find
\begin{eqnarray}
&& J_{tr}(\omega_1) \simeq -2\left(\ln \frac{\omega_1}{\omega_p}+1 \right),
\quad \int_{0}^{\omega'}\frac{d\omega_1}{d\omega_1}
\left(\ln \frac{\omega_1}{\omega_p}+1 \right)
\left[1-\exp\left(-ix\frac{\omega_1}{\omega}\right) \right] \nonumber \\
&& =\int_{0}^{b} \frac{dz}{z} \left(\ln \frac{\omega}{x\omega_p}+1+
\ln z \right) \left[1-\exp\left(-ix \right) \right], \quad
b=x\frac{\omega'}{\omega}
\label{32}
\end{eqnarray}
Since the probability $dw(\omega_1)/d\omega_1$ in (\ref{5})
is relatively small (this situation could changes drastically if one
considers a pile consisting of many plates) and the region $x \sim 1$
contributes mainly into the integral over $x$ we have $b \gg 1$,
one can expand integral in (\ref{32}) over $1/b$. Conserving the main
term of this expansion and adding the result to formula (\ref{31}) we
obtain
\begin{equation}
\int_{0}^{\infty} J_{tr}(\omega_1)
\left[1-\exp\left(-ix\frac{\omega_1}{\omega}\right) \right]
\frac{d\omega_1}{\omega_1} \simeq
\left(\ln x+\sigma \right)\left(\ln x+\sigma +i\pi\right),
\label{33}
\end{equation}
where
\begin{equation}
\sigma=\ln \frac{\omega_p}{\omega}+C-1
\label{34}
\end{equation}
Substituting the expression obtained into formula (\ref{5}) and
taking into account the factor $g(\beta)$ (\ref{a.15})
we have
\begin{equation}
\frac{d\varepsilon}{d\omega}=\frac{g(\beta)}{\pi} \exp(-w_c)~{\rm Re}~
\int_{0}^{\infty}\exp \left[ix-\eta
\left(\ln x+\sigma \right)\left(\ln x+\sigma +i\pi\right) \right]dx
\label{35}
\end{equation}
In the integral in (\ref{35}) we turn the integration contour by the angle
$\pi/2$. We find
\begin{eqnarray}
&& \frac{d\varepsilon}{d\omega}=\frac{g(\beta)}{\pi} \exp(-w_c)~{\rm Im}~
\exp \left[-\eta\left(\sigma^2+2i\pi \sigma
-\frac{3\pi^2}{4}\right) \right] \nonumber \\
&& \times \int_{0}^{\infty}\exp \left(-x-\eta
\ln^2 x \right) x^{-2\eta\left(\sigma+i\pi\right)} dx
\label{36}
\end{eqnarray}
Expanding here $\exp \left(-\eta \ln^2 x \right) \simeq 1-\eta \ln^2 x$
($\eta \ll 1$) we have integrals
\begin{eqnarray}
&& \int_{0}^{\infty}\exp \left(-x \right)
x^{-2\eta\left(\sigma+i\pi\right)} dx = \Gamma\left(1-2\eta(\sigma
+i\pi) \right), \nonumber \\
&& \int_{0}^{\infty}\exp \left(-x \right) \ln^2 x
x^{-2\eta\left(\sigma+i\pi\right)} dx = \Gamma''\left(1-2\eta(\sigma
+i\pi) \right),
\label{37}
\end{eqnarray}
where $\displaystyle{\Gamma''(z)=\frac{d^2\Gamma(z)}{dz^2}}$. Substituting
these integrals into (\ref{36}) and taking into account that
$\eta \ll 1$ (but it may be that $\eta \sigma \sim 1$) we obtain
\begin{eqnarray}
&& \frac{d\varepsilon}{d\omega}=\frac{g(\beta)}{\pi} \exp(-w_c)
\exp \left[-\eta\left(\sigma^2
-\frac{3\pi^2}{4}+\psi^2(1-2\eta \sigma)+\psi'(1-2\eta \sigma)
\right) \right] \nonumber \\
&& \times \sin \left[2\pi \eta \left(\sigma+\psi(1-2\eta \sigma)
\right)\right] \Gamma(1-2\eta \sigma),
\label{38}
\end{eqnarray}
where $\psi(z)=d \ln \Gamma(z)/dz$. In the case $\eta \sigma \ll 1$
we have from (\ref{38}), (\ref{34})
\begin{eqnarray}
&& \frac{d\varepsilon}{d\omega}=
2\eta \left(\ln \frac{\omega_p}{\omega}-1\right)g(\beta) \exp(-w_c) f_{tr},
\nonumber \\
&& f_{tr}=\exp \left[-\eta\left(\ln \frac{\omega_p}{\omega}-1\right)^2
\left(1+\frac{\eta \pi^2}{3} \right)+\frac{\eta \pi^2}{4} \right].
\label{39}
\end{eqnarray}

The case of a thin target, where the photon formation length 
\begin{equation}
l_f = \frac{2\gamma^2}{\omega}\left(\kappa + \nu_0 \right)
\geq l,
\label{40}
\end{equation}
$l$ is the target thickness, should be analyzed separately. We
consider situation when effects of the polarization of a medium are weak
($\kappa_0^2 < \nu_0$)
In this case the scattering takes place on a target as a whole
during the radiation process
and the spectral probability $dw/d\omega$ in Eqs.(\ref{2})-(\ref{5})
depends on the momentum transfer $q=\gamma \vartheta$ (see Sec.V of \cite{8},
$q$ is measured in the electron mass)
\begin{equation}
\frac{dw_r}{d\omega}=\frac{2\alpha}{\pi} \frac{d\omega}{\omega}
F_2\left(\frac{q}{2}\right) \equiv \beta_1(q)\frac{d\omega}{\omega},~
F_2(x)=\frac{2x^2+1}{x\sqrt{1+x^2}}\left(\ln \left(x+\sqrt{1+x^2} 
\right)-1 \right),
\label{41}
\end{equation}
where the function $F_2(x)$ is defined in Eq.(5.17) of \cite{8}.
On the final step one have to average $d\varepsilon_r/d\omega$ over
${\bf q}$ with the distribution function $F_s({\bf q})$ (see Eq.(5.18) of
\cite{8}).

Since the expression for the spectral distribution (\ref{41}) has
the same form as in the Bethe-Heitler case, one can use Eqs.(\ref{9}) and
(\ref{12}) for calculation of the reduction factor. As a result we
have for the contribution of the thin target region
\begin{equation}
f_{th}=\frac{1}{\left<\beta_1(q)\right>} \left<\beta_1(q)
\exp \left[-\beta_1(q) \ln \frac{\omega_{th}}{\omega}\right]\right>
\simeq 1-\frac{2\alpha}{\pi}
\displaystyle{\frac{\left<F_2^2\left(\frac{q}{2}\right) \right>}
{\left<F_2\left(\frac{q}{2}\right) \right>}\ln \frac{\omega_{th}}{\omega}},
\label{42}
\end{equation}
where $\omega_{th}$ is the boundary energy starting from which the target
becomes thin. The lower limit of the applicability of Eq.(\ref{42}) is
determined from the condition $l_f=l$. The expression (\ref{42})
depends essentially on the mean square of the momentum transfer $q_s^2$
associated with the target thickness. In the case $q_s^2 \ll 1$ when
an influence of the multiple scattering is weak
($\vartheta_s^2 \ll 1/\gamma^2$)
\begin{equation}
F_s(q)=\frac{4Z^2\alpha^2}{\left(q^2+\alpha_{s}^2 \right)^2}\frac{nl}{m^2},
\quad \alpha_s=\frac{\lambda_c}{a_{s2}},
\label{43}
\end{equation}
where $a_{s2}$ is the effective screening radius (see Eq.(2.19) in \cite{8}).
When the multiple scattering is strong ($q_s^2 \gg 1,~
\vartheta_s^2 \gg 1/\gamma^2$), the reduction factor can be calculated
using the Gaussian distribution for $F_s(q)$
\begin{equation}
F_s(q)=\frac{1}{\pi q_s^2} \exp \left(-\frac{q^2}{q_s^2} \right),~
q_s^2=\frac{4\pi Z^2\alpha^2}{m^2} nl \int_{0}^{q_s^2+1}
\frac{q^2dq^2}{\left(q^2+\alpha_{s}^2 \right)^2} \simeq
\frac{4\pi Z^2\alpha^2}{m^2} nl \ln \frac{q_s^2+1}{\alpha_s^2}.
\label{44}
\end{equation}
In the first case ($q_s^2 \ll 1$) one has $\left<F_2^2 \right> \ll
\left<F_2 \right>$ and one can neglect the correction to unit in (\ref{42}).
In the opposite case ($q_s^2 \gg 1$) the main contribution into entering
mean values of $F_2$ gives values $q \gg 1$ where
\begin{equation}
F_2(q) \simeq \ln q^2 - 1 + \frac{2}{q^2}.
\label{45}
\end{equation}
In this case the effect under consideration may be noticeable.

\section{Basic properties of the multi-photon radiation. Comparison with
experimental data}
\setcounter{equation}{0}

From the analysis in Sec. 2 one can find an approximate expression
for the reduction factor in the general form. One can present the integral
over $\omega_1$ in (\ref{5}) as
\begin{equation}
\int_{0}^{\infty}\frac{dw}{d\omega_1}
\left[\exp\left(-ix\frac{\omega_1}{\omega}\right)-1 \right] d\omega_1
\simeq -\int_{\omega}^{\infty}\frac{dw}{d\omega_1}d\omega_1
+\int_{0}^{\omega}\frac{dw}{d\omega_1}
\left[\exp\left(-ix\frac{\omega_1}{\omega}\right)-1 \right] d\omega_1.
\label{3.1}
\end{equation}
The second integral on the right-hand side of (\ref{3.1}) (as well as
terms omitted in the first integral) has an order $\omega dw/d\omega$.
For the case $\beta \ll 1$ this value is small
\begin{eqnarray}
&& \omega \frac{dw}{d\omega} \leq
\frac{4l}{3L_{rad}} \equiv \beta \quad (\omega \geq \omega_p);
\nonumber \\
&& \omega \frac{dw}{d\omega} \simeq \frac{4\alpha}{\pi}
\left(\ln \frac{\omega_p}{\omega}-1 \right)=2\eta(\sigma-C),\quad
\eta=\frac{2\alpha}{\pi} \quad (\omega \ll \omega_p).
\label{3.2}
\end{eqnarray}
Let us note that a contribution of the transition radiation can be
enlarged $n$ times if one makes a target as a collection of $n$ plates
conserving the total thickness $l$ provided that definite conditions are
fulfilled for the plate thicknesses and gaps between plates. In that
case $\eta \rightarrow n \eta$ and formulae of Sec.2 are valid if
$n \eta \leq 1$.

Expanding the exponential in (\ref{5}) with the second integral of
(\ref{3.1}) and integrating over $x$ we have
\begin{eqnarray}
&& \frac{1}{2\pi} \int_{-\infty}^{\infty}\exp (ix) dx
\int_{0}^{\omega}\frac{dw}{d\omega_1}
\left[\exp\left(-ix\frac{\omega_1}{\omega}\right)-1 \right] d\omega_1
\nonumber \\
&&=\int_{0}^{\omega}\frac{dw}{d\omega_1}\delta\left(1-\frac{\omega_1}
{\omega} \right)d\omega_1 = \omega \frac{dw}{d\omega}
\label{3.4}
\end{eqnarray}
Thus, the spectral distribution of the energy losses and reduction factor
$f$ have the following general form
\begin{equation}
\frac{d\varepsilon}{d\omega}=\omega \frac{dw}{d\omega} f,\quad
f= \exp \left[-\int_{\omega}^{\infty}\frac{dw}{d\omega_1}d\omega_1 \right]
\left(1+O\left(\omega \frac{dw}{d\omega} \right) \right)
\label{3.5}
\end{equation}

The cross section of the bremsstrahlung process
in the photon energies region where the influence of the LPM is very strong
was calculated in \cite{8} with term $\propto 1/L$ , where $L$
is characteristic logarithm of the problem,
and with the Coulomb corrections taken into account.
The polarization of a medium was included into
the approach.  We calculated also the contribution
of a photon emission on the boundaries of a target which involves
the transition radiation.
The energy losses spectrum in the tungsten target
with the thickness $l = 2.7\% L_{rad} (\beta = 0.036)$ for the initial
electron energies $\varepsilon$ = 25~GeV  and $\varepsilon$ = 8~GeV
was calculated and shown in Fig.2 in \cite{8}.
The contribution of an inelastic scattering of
a projectile on atomic electrons was not included into the
numerical calculation (it is $\sim 1~\%$), although it can be done
using Eq.(2.46) of \cite{8}. Here in Fig.3 we add to these results
the multi-photon effects. The curves 1-5 is the same curve as in Fig.2
of \cite{8}, the curve T involves the reduction factor which we
constructed as the interpolation of Eqs.(\ref{13}), (\ref{26}),
(\ref{39}) with accuracy up to $ 1~\%$.

The curves T in Fig.3 (a) and (b) are the final theory prediction in the
units $2\alpha/\pi$. Experimental data are taken from \cite{3} and
recalculated according with given in this paper procedure
\begin{equation}
\left(\frac{d\varepsilon}{d\omega}\right)_{exp}=
\frac{l}{L_{rad}} \frac{N_{exp}}{k}
\label{3.7}
\end{equation}
In our papers \cite{8} and \cite{8b} we put $k$=0.09 according with the
instruction given in paper \cite{7}. However, in the recent review \cite{9c}
it was definitely indicated that $k$=0.096 what is correspond to the
definition (photon energies were histogrammed logarithmically,
using 25 bins per decade of energy). So, in this paper we used
$k$=0.096. Data recalculated with this coefficient are also given
in Fig.3. It is seen that there is the perfect agreement of
the curves T with data for both energies.

In our paper \cite{8b} we compared calculation for the gold target
with the thickness $l=0.7~\%~L_{rad}$ and energy $\varepsilon$=25~GeV
with data from \cite{7}. In this case the reduction factor $f \simeq 0.94$
for photons with energy $\omega < 10~$ MeV (plateau region).
From the other side, use of the coefficient $k$ = 0.096 in (\ref{3.7})
instead of $k$ = 0.09 lowers data upon $6\%$. As a result, the excellent
agreement of the theory and data marked in \cite{8b} is not broken.

It is seen in Fig.3, as well as in Fig.2 \cite{8} and in Fig.1
\cite{8b}, that in the soft part of the measured spectrum the
boundary photons emission dominates. The formulae describing this effect
were derived in \cite{8} and \cite{8b} and this is
allowed to obtain description
of the measured spectrum on the whole. The authors of \cite{7} came across
some problems with surface effects theory. Because of this an
attempt was made to eliminate the boundary
radiation by subtracting spectra from targets made of the same
material, but different thicknesses (e.g. 5~\% - 3~\% of $L_{rad}$
in uranium (U)) assuming that an interference between the two
target edges is negligible (this is true for thick target \cite{8b}).
However, the subtraction gives no direct quantitative information
about boundary radiation because the effect of the multiple photon emission
is rather strong (more than 30~\% in the target with the thickness
5~\% $L_{rad}$). One can conclude
only that this difference should vanish (if thicknesses are not very
different) in the region where the contribution
of the boundary radiation is comparable
with radiation inside a target. In the photon energy interval where
the transition radiation dominates this difference is negative and
it is proportional to the difference of the reduction factors.
There is no data in the mentioned interval.

We would like to thank S. Klein who attract our attention to important
role of many photons emission at the SLAC experiment.

\newpage
\setcounter{equation}{0}
\appendix

\section{Appendix}

In the case when the spectral distribution of the probability of radiation
has the form
\begin{equation}
\frac{dw}{dx}=\frac{\beta}{x}f(x),\quad x=\frac{\omega}{\varepsilon},
\quad f(0)=1,
\label{a.1}
\end{equation}
the probability of sequential emission of $n$ photons
with energies $\omega_k \geq \omega_b
~(k=1 \ldots n)$ taking into account a recoil is defined by the
$n$-fold integral
\begin{equation}
F_n==\frac{1}{n!}\int_{\Delta}^{1}\frac{dx_1}{x_1}f(x_1)
\int_{\Delta_1}^{1}\frac{dx_2}{x_2}f(x_2) \ldots
\int_{\Delta_{n-1}}^{1}\frac{dx_n}{x_n}f(x_n),
\label{a.2}
\end{equation}
where
\begin{equation}
\Delta_k=\frac{\Delta}{1-y_k},\quad y_k=\sum_{m=1}^{k}x_m,\quad
\Delta=\frac{\omega_b}{\varepsilon}.
\label{a.3}
\end{equation}
Here $\omega_b$ is the lower boundary of photon energy which we consider,
we assume that this energy is small comparing with the particle energy
$\omega_b \ll \varepsilon$. Below we neglect terms order of $\Delta$,
then the function $F_n$ depends on $\ln (1/\Delta)$ only. Because of this
\begin{eqnarray}
&& F_n(\ln \frac{1}{\Delta})=\frac{1}{n}
\int_{\Delta}^{1}\frac{dx}{x}\left[1+xf_1(x) \right]
F_{n-1}\left(\ln \frac{1-x}{\Delta} \right) \nonumber \\
&& =\frac{1}{n}\int_{0}^{1-\Delta}
dz\left[\frac{1}{1-z}+f_2(z) \right]F_{n-1}\left(\ln \frac{1}{\Delta}+\ln z
\right),
\label{a.4}
\end{eqnarray}
where $f(x)=1+xf_1(x),~ f_2(z)=f_1(1-z)$.\\
We assume that $f_2(z)$ is a polynomial of degree $m$
\begin{equation}
f_2(z)=\nu_1+\nu_2z+\ldots \nu_mz^{m-1}
\label{a.5}
\end{equation}
Using the integrals
\begin{eqnarray}
&& \int_{0}^{1-\Delta}\frac{dz}{1-z}=\ln \frac{1}{\Delta},\quad
\int_{0}^{1-\Delta}\frac{dz}{1-z} \ln^n z \simeq
\int_{0}^{1}\frac{dz}{1-z} \ln^n z = (-1)^n n!~\zeta (n+1),\nonumber \\
&& \int_{0}^{1} dz z^m \ln^n z = (-1)^n \frac{n!}{(m+1)^{n+1}},
\label{a.6}
\end{eqnarray}
we find the following recursive relation
\begin{eqnarray}
&&\hspace{-10mm} F_n=\frac{1}{n}\sum_{k=0}^{n-1}(-1)^k s_{k+1}F_{n-k-1},\quad
F_0=1, \nonumber \\
&&\hspace{-10mm}s_n=a_n+\nu_1+\frac{\nu_2}{2^n}+\frac{\nu_3}{3^n}
+\ldots \frac{\nu_m}{m^n},
\quad a_1=\ln \frac{1}{\Delta},\quad a_n=\zeta(n)~(n \geq 2) .
\label{a.7}
\end{eqnarray}
One can show that the recursive relation (\ref{a.7}) may be obtained by
rearrangement of the Taylor series of the exponential function
(compare e.g. with the series for $\Gamma(1+\beta)$ \cite{10}, Sec.8.32)
\begin{equation}
F(\beta)=\exp \left[\sum_{n=1}^{\infty}\frac{(-1)^{n-1}}{n}s_n \beta^n
\right]=\sum_{n=0}^{\infty}F_n \beta^n.
\label{a.8}
\end{equation}
Using the explicit form of the coefficients $s_n$ (\ref{a.7}) we find
\begin{eqnarray}
&& \hspace{-10mm}F(\beta)=\exp \bigg[\beta\left(\ln \frac{1}{\Delta}-C \right)
-\ln \Gamma(1+\beta)+\nu_1 \ln (1+\beta)+\nu_2 \ln \left(1+\frac{\beta}{2}
\right) \nonumber \\
&& \hspace{-10mm}+ \ldots \nu_m \ln \left(1+\frac{\beta}{m} \right) \bigg]
=\frac{1}{\Gamma(1+\beta)}\exp \left[\beta
\left(\ln \frac{1}{\Delta}-C \right) \right]
\prod_{k=1}^{m}\left(1+\frac{\beta}{k} \right)^{\nu_k}
\label{a.9}
\end{eqnarray}
For the Bethe-Heitler formula (\ref{7}) one has
\begin{equation}
\nu_1=-\frac{1}{4},\quad \nu_2=-\frac{3}{4},\quad \nu_m=0~(m \geq 3),
\label{a.10}
\end{equation}
so that
\begin{equation}
F_{BH}(\beta)=\exp \left[\beta\left(\ln \frac{1}{\Delta}-C \right) \right]
\frac{1}{\Gamma(1+\beta)(1+\beta)^{1/4}\left(1+\beta/2 \right)^{3/4}}.
\label{a.11}
\end{equation}
For radiation of the scalar particle one has $\nu_1=-1,~\nu_m=0~(m \geq 2)$
so that
\begin{equation}
F_{s}(\beta)=\exp \left[\beta\left(\ln \frac{1}{\Delta}-C \right) \right]
\frac{1}{\Gamma(1+\beta)(1+\beta)}.
\label{a.12}
\end{equation}
So the value $A$ in Eq.(\ref{3}) has the form
\begin{equation}
A(\beta)=F_{BH}^{-1}(\beta)=\exp \left[-\beta\left(\ln \frac{1}{\Delta}-C
\right) \right]
\Gamma(1+\beta)(1+\beta)^{1/4}\left(1+\beta/2 \right)^{3/4}.
\label{a.13}
\end{equation}
Just the normalization factor $A(\beta)$ (this is the probability 
of the passage of a projectile through a target without 
radiation) should enter in Eq.(\ref{5})
instead of $\exp (-w)$. Thus, when we multiply the right-hand side of
(\ref{5}) by the factor
\[
g(\beta)=\exp \left(w)\right)A(\beta),
\]
we take into account accurately the emission of arbitrary number of the
hard photons. This extends Eq.(\ref{5}) on the case $\beta \geq 1$ if
the radiation of the hard ($\omega \sim \varepsilon$) photons is
given by the Bethe-Heitler formula.

Taking into account that the probability of radiation of the photon with
the energy larger than $\omega_b$ is (see (\ref{8}))
\begin{equation}
w(\Delta)=\beta\left(\ln \frac{1}{\Delta}-\frac{5}{8} \right),
\label{a.14}
\end{equation}
we obtain following expression for the factor $g(\beta)$
\begin{equation}
g(\beta)=\exp \left(w(\Delta)\right)A(\beta)
=\exp \left[\beta\left(C-\frac{5}{8}\right) \right]
\Gamma(1+\beta)(1+\beta)^{1/4}\left(1+\beta/2 \right)^{3/4}.
\label{a.15}
\end{equation}
As one expected, the function $g(\beta)$ is independent of the boundary
energy $\omega_b~(\Delta)$ and its expansion at $\beta \ll 1$ has
the form $g(b)= 1+ O(\beta^2)$.
The function $g(\beta)$ is shown in Fig.4. One can see that
$g(\beta) \simeq 1$ for $\beta < 0.2$, a noticeable growth begins at
$\beta \sim 0.5$, and at $\beta \sim 1$ the function $g(\beta)$ increases
very fast.

This factor enters into the generalization of formula (\ref{5})
for the case of a thick target ($\beta \geq 1$)
\begin{equation}
\frac{d\varepsilon}{d\omega}==\frac{g(\beta)}{\pi}~{\rm Re}
\int_{0}^{\infty}\exp\left(ix \right)
\exp\left\{-\int_{0}^{\infty}\frac{dw}{d\omega_1}
\left[1-\exp\left(-ix\frac{\omega_1}{\omega}\right)\right]
d\omega_1\right\}dx
\label{a.16}
\end{equation}
It should be noted that Eq.(\ref{5}) was obtained as the solution of 
the kinetic equation for the case when the probability of radiation
$dw/d\omega$ is independent of the particle energy $\varepsilon$. After
the factor $g(\beta)$ is incorporated, Eq.(\ref{a.16}) is the solution of 
the corresponding kinetic equation in the soft part of the energy
losses ($(1+\beta)\omega \ll \varepsilon$) in the case when the mentioned
dependence of the particle energy $\varepsilon$ is included.

\newpage

{\bf Figure captions}

\vspace{15mm}
\begin{itemize}

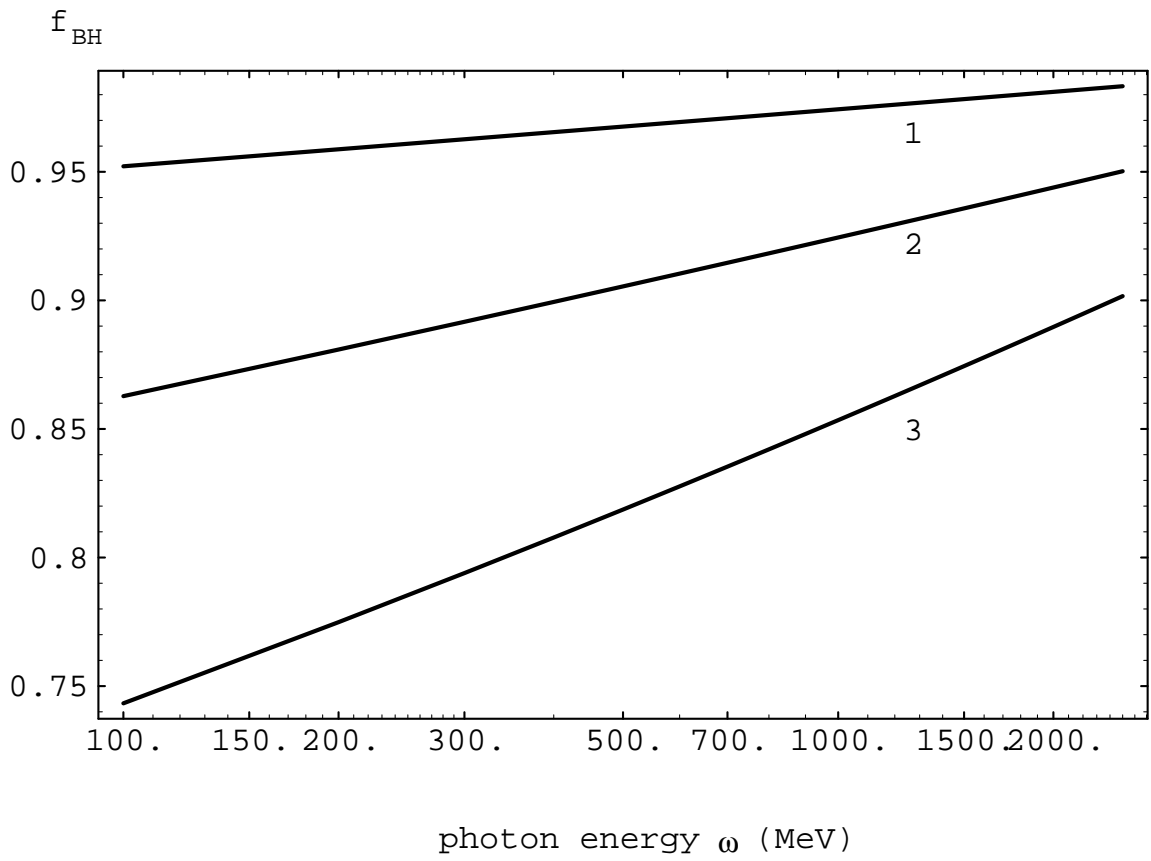
\item {\bf Fig.1} The reduction factor $f_{BH}$ for energy of the initial
electron $\varepsilon$=25~GeV. The curves 1, 2 and 3 are respectively for
$\beta$=0.01, 0.03 and 0.06.

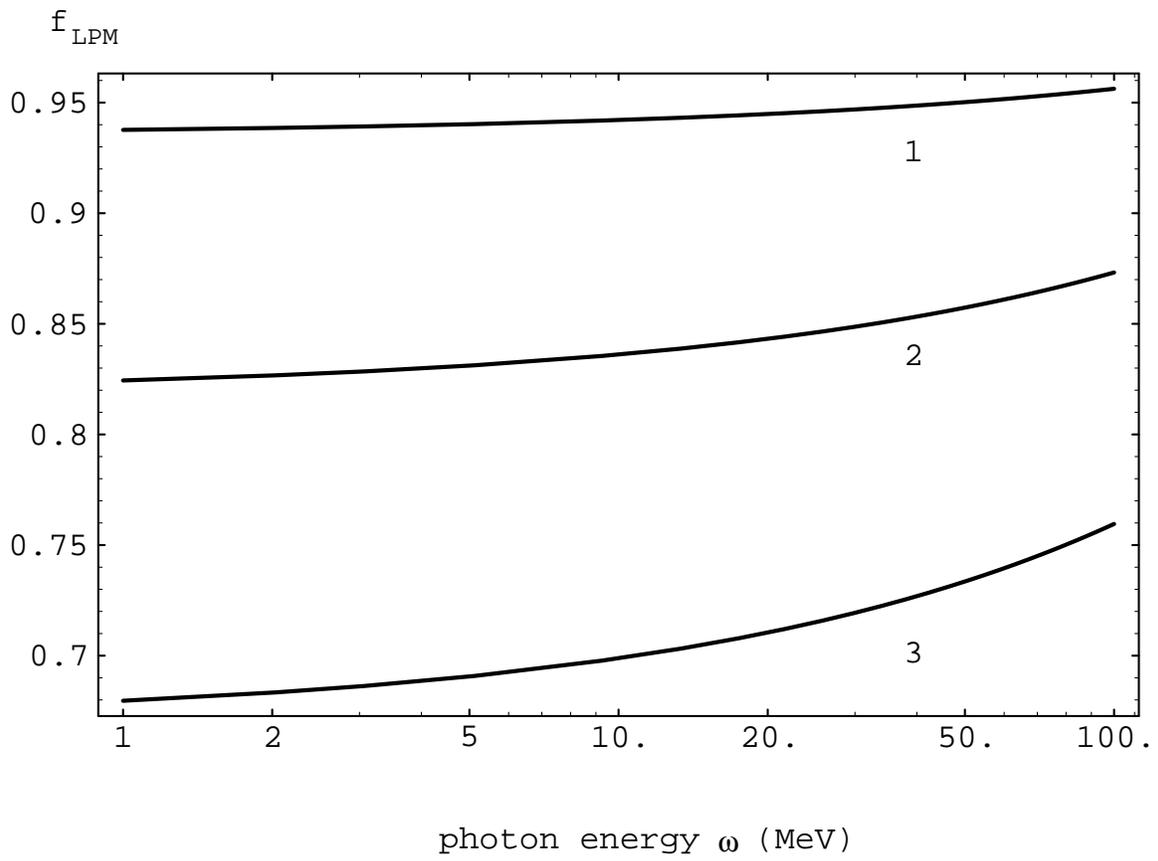
\item {\bf Fig.2} The reduction factor $f_{LPM}$ for energy of the initial
electron $\varepsilon$=25~GeV in tungsten ($\omega_c$=228~MeV).
The curves 1, 2 and 3 are respectively for
$\beta$=0.01, 0.03 and 0.06.

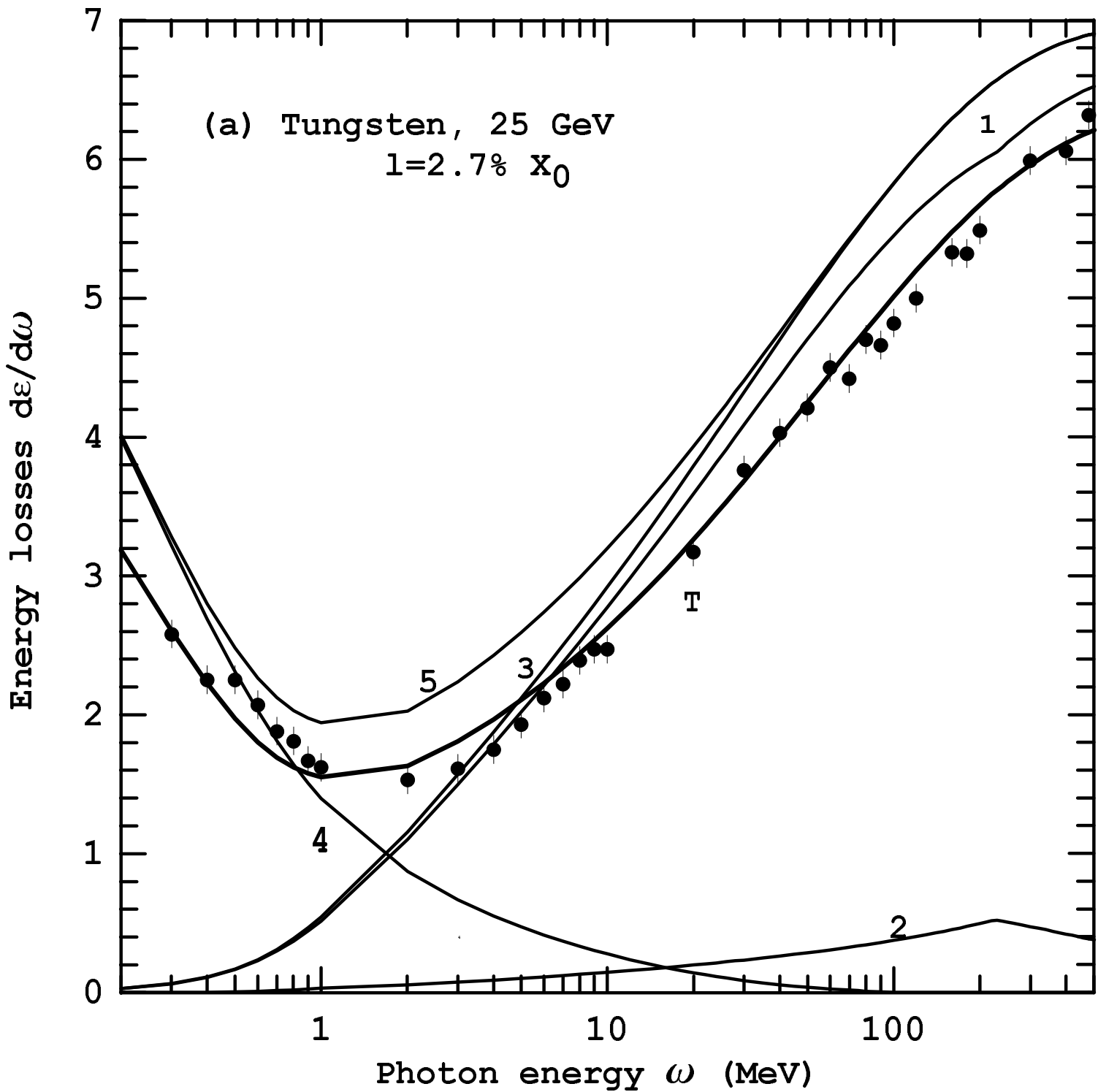
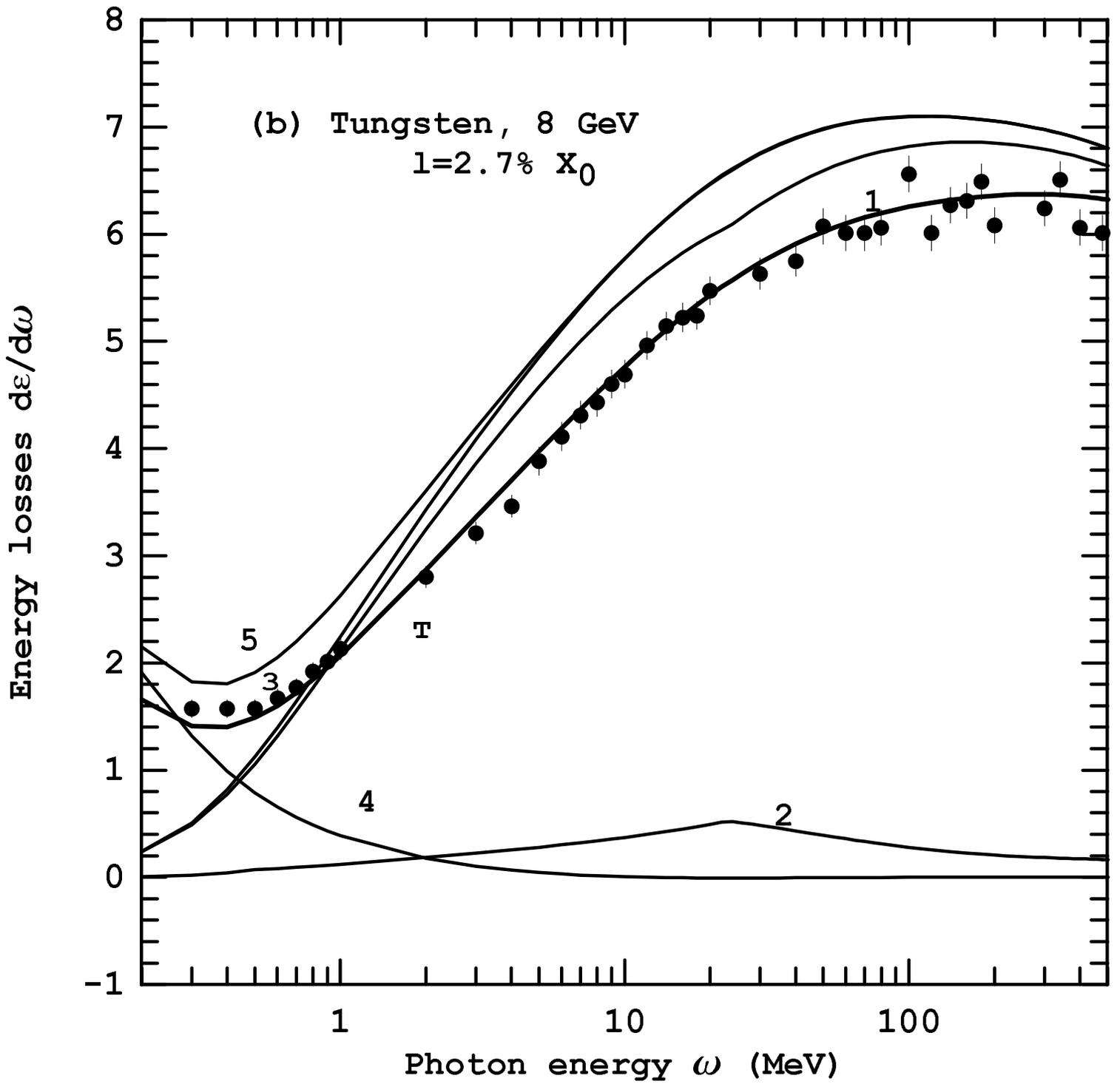
\item {\bf Fig.3} The energy losses
$\displaystyle{\frac{d\varepsilon}{d\omega}}$
in tungsten with thickness $l=0.088~mm$ in units
$\displaystyle{\frac{2\alpha}{\pi}}$,
((a) is for the initial electrons energy $\varepsilon=25~GeV$ and (b)
is for $\varepsilon=8~GeV$).
The Coulomb corrections and the polarization of a medium are included.
\begin{itemize}
\item Curve 1 is the contribution of the main term describing LPM effect;
\item curve 2 is the correction term;
\item curve 3 is the sum of the previous contributions;
\item curve 4 is the contribution of the boundary photons;
\item curve 5 is the sum of the previous contributions;
\item curve T is the final theory prediction with regard for
the reduction factor (the multiphoton effects).
\end{itemize}
Experimental data from Fig.9  of \cite{7}.

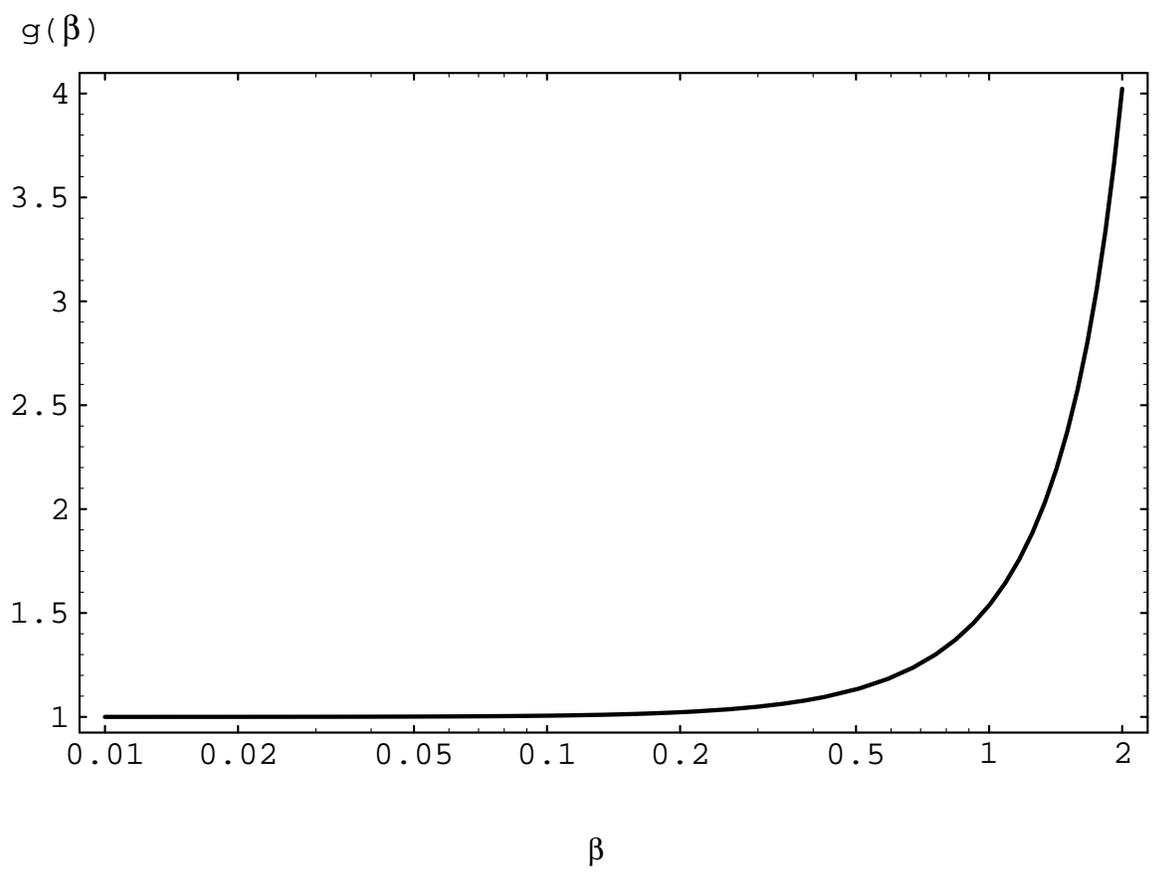
\item {\bf Fig.4} The function $g(\beta)$ Eq.(\ref{a.15}).
\end{itemize}


\newpage

\begin{thebibliography}{99}
\bibitem{5} P. L. Anthony, R. Becker-Szendy, P. E. Bosted {\em et al},
Phys. Rev. Lett. {\bf 75} (1995) 1949.
\bibitem{6} P. L. Anthony, R. Becker-Szendy, P. E. Bosted {\em et al},
Phys. Rev. Lett. {\bf 76} (1996) 3550.
\bibitem{7} P. L. Anthony, R. Becker-Szendy, P. E. Bosted {\em et al},
Phys.Rev. {\bf D56} (1997) 1373.
\bibitem{1} L. D. Landau and I. Ya. Pomeranchuk, Dokl.Akad.Nauk SSSR
{\bf 92} (1953) 535, 735. See in English in {\em The
Collected Paper of L. D. Landau}, Pergamon Press, 1965.
\bibitem{2} A. B. Migdal, Phys. Rev. {\bf 103} (1956) 1811.
\bibitem{3} A. B. Migdal, Sov. Phys. JETP {\bf 5} (1957) 527.
\bibitem{4} M. L. Ter-Mikaelian, {\em High Energy Electromagnetic
Processes in Condensed Media}, John
Wiley \& Sons, 1972.
\bibitem{8} V. N. Baier and V. M. Katkov,
Phys.Rev. D{\bf 57} (1998) 3146
\bibitem{8b} V. N. Baier and V. M. Katkov, {\em The Landau-Pomeranchuk-Migdal
effect in a thin target}, hep-ph 9712524,
Preprint BINP 97-105, Novosibirsk, 1997.
\bibitem{9a} L. D. Landau J.Phys. USSR {\bf 8} (1944) 201.
\bibitem{9b} J. D. Jackson {\em Classical Electrodynamics} 2nd ed.
Wiley \& Sons, 1975.
\bibitem{9c} S. Klein {\em Suppression of Bremsstrahlung and Pair Production
due to Environmental Factors}, hep-ph 9802442, LBL-41350, Berkeley, 1998.\\
(submitted to Rev. Mod. Phys.)
\bibitem{10} I. S. Gradshtein and I. M. Ryzhik,
{\em Table of Integrals, Series, and Products},
Academic Press, New York, 1965.

\end{thebibliography}
\end{document}